\documentclass[12pt]{article}
\usepackage{epsfig}
\usepackage{graphicx}
\usepackage{amsmath}
\usepackage{amssymb}
\usepackage{slashed}
\usepackage{epstopdf}

\newcommand{\beq}{\begin{equation}}
\newcommand{\eeq}{\end{equation}}
\newcommand{\bea}{\begin{eqnarray}}
\newcommand{\eea}{\end{eqnarray}}
\newcommand{\bem}{\begin{multline}}
\newcommand{\eem}{\end{multline}}
\newcommand{\beg}{\begin{gather}}
\newcommand{\eeg}{\end{gather}}

\def\eq#1{{Eq.~(\ref{#1})}}

\newcommand{\ben}{\begin{eqnarray*}}
\newcommand{\een}{\end{eqnarray*}}

\begin{document}
\title{{\bf A model for gluon production in heavy-ion collisions at
    the LHC with rcBK unintegrated gluon densities \\[0cm] }}
\author{
{\bf Javier L. Albacete$^a$\thanks{e-mail:
    javier.lopez-albacete@cea.fr},\hspace{0.1cm}
 Adrian Dumitru$^b$\thanks{Adrian.Dumitru@baruch.cuny.edu}}
\\[0.15cm] {\it \small $^a$Institut de Physique Th\'eorique -
  CEA/Saclay,  91191 Gif-sur-Yvette cedex, France.}\\
 {\it \small $^b$Department of Natural Sciences, Baruch College, CUNY,
17 Lexington Avenue, New York, NY 10010, USA}\\
{\it \small RIKEN BNL Research Center, Brookhaven National
Laboratory, Upton, NY 11973, USA }\\
}
\maketitle

\begin{abstract}
This note is a physics manual for a recent numerical implementation of
$k_t$-factorization with running-coupling BK unintegrated gluon
distributions. We also compile some results for Pb+Pb collisions at
$\surd s = 2.75$~TeV, such as predictions for the centrality
dependence of the charged particle multiplicity, transverse energy,
and eccentricity. The model can further be used to obtain initial
conditions for hydrodynamic simulations of A+A collisions at the LHC.
\end{abstract}

\section{Introduction}

In this note we describe some of the physics underlying a recent
numerical implementation of $k_t$-factorization with running-coupling
BK unintegrated gluon distributions for heavy-ion (and pp, pA) collisions
at the RHIC and LHC colliders. Its main purpose is to describe the
framework as well as some of the main assumptions and model parameters
which could be tested by comparing to upcoming data from heavy-ion
collisions at the LHC.

Further, the model may be useful for obtaining initial conditions for
hydrodynamic simulations of A+A collisions. Below, we also summarize a
few key results. The C++ code is available for download at\\ 
\verb*#http://physics.baruch.cuny.edu/node/people/adumitru/res_cgc#.

In this ``manual'' we do not provide an overview of the theory or
applications of the Color Glass Condensate.

\section{The running coupling BK equation}

The Color Glass Condensate is equipped with a set of renormalization
group equations, the B-JIMWLK equations, which describe the quantum
evolution of hadron structure towards small-$x$ (see e.g. 
\cite{Gribov:1984tu, Iancu:2003xm,Weigert:2005us} and references therein). The
B-JIMWLK equations are equivalent to an infinite set of coupled
non-linear integro-differential evolution equations for the
expectation values of the different correlators of Wilson lines
averaged over the target gluon field configurations. In the
large-$N_c$ limit the full B-JIMWLK hierarchy reduces to the
Balitsky-Kovchegov (BK) \cite{Balitsky:1996ub,Kovchegov:1999yj}
equation: a single, closed equation for the the forward scattering
amplitude of a $q\bar{q}$ dipole on a dense target
\begin{equation}
\mathcal{N}(\underline{x},\underline{y},Y)=1-\frac{1}{N_c}\langle
U(\underline{x}) \, U^{\dagger}(\underline{y})\rangle_Y
\end{equation}
where the $U$'s denote Wilson lines in the fundamental representation,
$\underline{x}$ ($\underline{y}$) is the transverse position of the
quark (antiquark) and $\underline{r}=\underline{x}-\underline{y}$ the
dipole transverse size. The average over the target gluon field
configurations is performed at evolution rapidity $Y=\ln (x_0/x)$,
where $x_0$ is the starting point for the evolution. The BK equation
reads:
\begin{equation}
  \frac{\partial\mathcal{N}(r,x)}{\partial\ln(x_0/x)}=\int d^2{\underline r_1}\
  K({\underline r},{\underline r_1},{\underline r_2})
  \left[\mathcal{N}(r_1,x)+\mathcal{N}(r_2,x)
-\mathcal{N}(r,x)-\mathcal{N}(r_1,x)\,\mathcal{N}(r_2,x)\right]\,
\label{bk1}
\end{equation}
where $K$ is the evolution kernel and
$\underline{r_2}=\underline{r}-\underline{r_1}$. In \eq{bk1}
we have implicitly assumed translational invariance over scales of
order the nucleon radius, i.e.\ that the dipole amplitude depends only
on the dipole transverse size $r \equiv| \underline{r}|$ but not on
the impact parameter $\underline{b}=(\underline{x}+\underline{y})/2$
of the collision\footnote{This is the main reason why we presently
  refrain from applying the model to compute the energy dependence of
  the multiplicity in pp collisions; accurate results
  require this input, see for example~\cite{Tribedy:2010ab}.}. A smooth
variation of $\mathcal{N}(r,x)$ on larger distance scales can be
incorporated via the initial condition at the starting point $x_0$ of
the evolution, see below.

The original BK equation resums small-$x$ gluon emission to all orders
at leading-logarithmic (LL) accuracy in $\alpha_s\ln(x_0/x)$, with
$\alpha_s$ fixed, and also contains non-linear terms that account for
gluon-gluon self-interactions, relevant in a high density
scenario. However, such limited dynamical input does not suffice to
provide a good description of experimental data.  This situation has
been considerably improved by the recent determination of running
corrections to the LL equations
\cite{Kovchegov:2006vj,Balitsky:2006wa, Gardi:2006rp}. Such
corrections amount to just a modification of the evolution kernel in
\eq{bk1} with respect to the LL result. Though there are different
possibilities of defining the running coupling kernel, it was shown in
\cite{Albacete:2007yr} that the prescription proposed by Balitsky in
\cite{Balitsky:2006wa} minimizes the role of additional {\it
  conformal} corrections that arise at the same order as the running
coupling, making it better suited for phenomenological
applications. The corresponding running coupling kernel reads
\begin{equation}
  K^{{\rm run}}({\bf r},{\bf r_1},{\bf r_2})=\frac{N_c\,\alpha_s(r^2)}{2\pi^2}
  \left[\frac{1}{r_1^2}\left(\frac{\alpha_s(r_1^2)}{\alpha_s(r_2^2)}-1\right)+
    \frac{r^2}{r_1^2\,r_2^2}+\frac{1}{r_2^2}\left(\frac{\alpha_s(r_2^2)}{\alpha_s(r_1^2)}-1\right) \right]
\label{kbal}
\end{equation}
We shall refer to \eq{bk1} together with the evolution kernel
\eq{kbal} as the running coupling BK (rcBK) equation. Running coupling
corrections have proven essential for promoting the BK equation
to a phenomenological tool. Indeed, the rcBK equation has been
employed successfully to describe inclusive structure functions in e+p
scattering \cite{Albacete:2009fh,Albacete:2010sy}, the energy and rapidity
dependence of hadron multiplicities in Au+Au collisions at RHIC
\cite{Albacete:2007sm}, as well as single inclusive spectra in p+p and
d+Au collisions at RHIC \cite{Albacete:2010bs}.
A detailed discussion on the numerical set up employed to solve the rcBK
equation can be found in \cite{Albacete:2007yr}. The present model
differs from earlier predictions published in
ref.~\cite{Albacete:2007sm} mainly due to the initial conditions
described below.

\eq{bk1} needs to be suplemented with initial conditions. We consider two different families of initial conditions labeled by the parameter $\gamma$:
\begin{equation}
\mathcal{N}(r,Y\!=\!0; R)=
1-\exp\left[-\frac{\left(r^2\,Q_{s0}^2(R)\right)^{\gamma}}{4}\,
  \ln\left(\frac{1}{\Lambda\,r}+e\right)\right]\ ,
\label{mv}
\end{equation}
where $\Lambda=0.241$ GeV. For each of these families there are two  free parameters: the
value $x_0$ where the evolution starts, which we fix it to $x_0=0.01$ in all cases, and the initial saturation
scale $Q_{s0}(R)$ at the transverse coordinate $R$. It measures the
local density of large-$x$ sources at a fixed point in impact
parameter space (i.e., in the transverse plane).
The case $\gamma=1$ corresponds to the McLerran-Venugopalan model. For a single nucleon we take the value $Q_{s0}^2=0.2$ GeV$^2$ for it and we refer to it in the plots below as MV ic. However, global fits to structure functions in electron+proton scattering at small-$x$ based on rcBK dynamics \cite{Albacete:2009fh,Albacete:2010sy} clearly indicate that a value $\gamma>1$ is preferred by the data. Following the results in  \cite{Albacete:2010sy}\footnote{Different parameter sets are obtained in \cite{Albacete:2010sy}. The values ($Q_{s0}^2,\gamma$) considered here correspond to fit h') in Table 1, which provide a best fit to data $\chi^2/d.o.f=1.104$. }, we take the values $Q_{s0}^2=0.168$ GeV$^2$ (for a single nucleon) and $\gamma=1.119$. We refer to this other set of initial conditions as MV$^{\gamma=1.119}$. The value of $\gamma$ determines the steepness of the unintegrated gluon distribution for momenta above the saturation scale $k_t>Q_s$, the MV$^{\gamma=1.119}$ initial conditions yielding a steeper falloff than the MV ones for all the evolution rapidities relevant at RHIC and the LHC (see fig. 1 bottom). The main novelty in this updated version of the manual is the introduction of this new set of initial conditions with $\gamma>1$.

As explained in more detail below, the geometry of a given A+A
collision is determined by the fluctuations in the positions of the
nucleons in the transverse plane. Each configuration defines a
different local density in the transverse plane of each
nucleus. Obviously, the smallest non-zero local density corresponds to
the presence of a single nucleon. On the
other hand, in A+A collisions rare fluctuations can result in
collisions of a large number of nucleons at the same transverse
position and, therefore, in a large $Q_{s0}$.  To account for all
possible configurations we tabulate the solution of the rcBK equation
for different values of the initial local density, i.e., for each value of
$Q_{s0}(R)$ in \eq{mv} ranging from $Q_{s0}^2=0.2$ (0.168) GeV$^2$ for MV (MV$^{\gamma=1.119}$) initial conditions (for a single nucleon) to $n\times Q_{s0}^2$ with $n=1\dots 30$. The solutions are then used in the $k_t$-factorization
formula to calculate local gluon production at each point in the
collision zone. Finally we perform the average over all the nucleon
configurations generated by the Monte Carlo.

To complete our discussion of the initial conditions we explain how we
construct $Q_{s0}({\bf R})$. We first generate a configuration of nucleons
for each of the colliding nuclei. This consists of a list of random
coordinates ${\bf r}_i$, $i=1\dots A$, chosen from a Woods-Saxon
distribution. Multi-nucleon correlations are neglected except for
imposing a short-distance hard core repulsion which enforces a minimal
distance $\approx 0.4$~fm between any two nucleons. After this step,
the longitudinal coordinate of any nucleon is discarded, they are
projected onto the transverse plane. Factorizing the fluctuations of
the nucleons in a nucleus from possible fluctuations of large-$x$
``hot spots'' within a nucleon (not accounted for at present), and
finally from semi-hard gluon production appears to be justified by the
scale hierarchy
\begin{equation}
\frac{1}{Q_s} \ll R_N \ll R_A ~,
\end{equation}
where $R_A$, $R_N$ are the radii of a nucleus and of a proton, respectively.

For a given configuration, the initial saturation momentum
$Q_{s0}({\bf R})$ at the transverse coordinate ${\bf R}$ is taken to
be
\begin{equation}
Q_{s0}^2({\bf R}) = N({\bf R}) \, Q_{s0,\, {\rm nucl}}^2~,
\end{equation}
where $Q_{s0,\, {\rm nucl}}^2=0.2$~GeV$^2$ for our original uGD with
MV model initial conditions while $Q_{s0,\, {\rm nucl}}^2=0.168$~GeV$^2$
for the new uGD with MV$^{\gamma=1.119}$ initial condition, as discussed
above.
$N({\bf R})$ is the number of nucleons from the given nucleus which
``overlap'' the point ${\bf R}$:
\begin{equation}
N({\bf R}) = \sum\limits_{i=1}^{A} \Theta \left(
\sqrt{\frac{\sigma_0}{\pi}} -  |{\bf R-r_i}|\right)~.
\end{equation}
Some care must be exercised in choosing the transverse area $\sigma_0$
of the large-$x$ partons of a nucleon. $Q_{s0}$ corresponds to the
density of large-$x$ sources with $x>x_0$ and should therefore be
energy independent (recoil of the sources is neglected in the
small-$x$ approximation). We therefore take $\sigma_0 \simeq 42$~mb to
be given by the inelastic cross-section at $\surd s =
200$~GeV. However, $\sigma_0$ should not be confused with the energy
dependent inelastic cross section $\sigma_{\rm in}(s)$ of a nucleon
which grows due to the emission of small-$x$ gluons.

\section{$k_t$-factorization}
According to the $k_t$-factorization formalism
\cite{Kovchegov:2001sc}, the number of gluons produced per unit
rapidity at a transverse position ${\bf R}$ in A+B collisions is given by

\begin{equation}
\frac{dN^{A+B\to g}}{dy\, d^2p_t\, d^2R} = \frac{1}{\sigma_s}
\frac{d\sigma^{A+B\to g}}{dy\, d^2p_t\, d^2R}\,,
\label{kt}
\end{equation}
where $\sigma_s$ represents the effective interaction area and
$\sigma^{A+B\to g}$ is the cross section for inclusive gluon
production:
\begin{equation}
\frac{d\sigma^{A+B\to g}}{dy\, d^2p_t\, d^2R} = K\,  \frac{2}{C_F} 
\frac{1}{p_t^2} \int^{p_t} \frac{d^2k_t}{4}\int d^2b\, 
\alpha_s(Q)\,
\varphi(\frac{|p_t+k_t|}{2},x_1;b)\,
\varphi(\frac{|p_t-k_t|}{2},x_2;R-b)~,
\label{kt2}
\end{equation}
with $x_{1(2)}=(p_t/\sqrt{s_{NN}})\exp(\pm y)$ and
$C_F=(N_c^2-1)/2N_c$. As noted before, we assume that the local
density in each nucleus is homogenous over transverse distances of the
order of the nucleon radius $R_N$. Thus, the $b$-integral in \eq{kt2}
yields a geometric factor proportional to the transverse ``area'' of a
nucleon which cancels with a similar factor implicit in $\sigma_s$
from \eq{kt}, modulo subtleties in the definition of $\sigma_s$.

The unintegrated gluon distributions (ugd's) $\varphi$ entering
\eq{kt2} are related to the dipole scattering amplitude in the adjoint
representation, $\mathcal{N}_G$, through a Fourier transform (for
consistency with the notation used in \eq{kt2} we make the impact
parameter dependence of the ugd's explicit):
\begin{equation}
\varphi(k,x,b)=\frac{C_F}{\alpha_s(k)\,(2\pi)^3}\int d^2{\bf r}\
e^{-i{\bf k}\cdot{\bf r}}\,\nabla^2_{\bf r}\,\mathcal{N}_G(r,Y\!=\!\ln(x_0/x),b)\,.
\label{phi}
\end{equation}
In turn, $\mathcal{N}_G$ is related to the quark dipole scattering
amplitude that solves the rcBK equation, $\mathcal{N}$, as follows:
\begin{equation}
\mathcal{N}_G(r,x)=2\,\mathcal{N}(r,x)-\mathcal{N}^2(r,x)\,.
\end{equation}
Note that this relation entails that the saturation momentum relevant
for gluon scattering is larger than that for quark scattering by about
a factor of 2.

Eqs.~(\ref{phi}) and (\ref{kt2}) were written originally for fixed
coupling. In order to be consistent with our treatment of the
small-$x$ evolution, we have extended them by allowing the coupling to
run with the momentum scale. The argument of the running coupling in
\eq{kt2} is chosen to be $Q={\rm max}\{|p_t+k_t|/2,|p_t-k_t|/2\}$,
while for the definition of the ugd \eq{phi} we take it to be the
transverse momentum itself, $k$. This turns out to be important in
order to reproduce the centrality dependence of charged particle
multiplicities at RHIC, which are otherwise too flat for small $N_{\rm
  part}$. However, the results are not very sensitive to the
particular choice of scale because $\varphi\to0$ as $k^2\to0$ due to
the saturation of $\mathcal{N}(r)$ at large dipole sizes $r$. In
principle, one could improve on this educated ansatz by using the
results of \cite{Horowitz:2010yg} where running coupling corrections
to inclusive gluon production have been studied. Most importantly, the
$x$-dependence of the dipole scattering amplitude obtained by solving
the rcBK equation encodes all the collision energy and rapidity
dependence of the gluon production formula \eq{kt2}.

The normalization factor $K\simeq2$ introduced in the
$k_t$-factorization formula~(\ref{kt2}) above is fixed by the charged
particle transverse momentum distribution in p+p collisions at 7~TeV,
see below. It lumps together higher-order corrections, sea-quark
contributions, a nucleon geometry factor, and so on. We apply an
additional ``gluon multiplication factor'' $\kappa_g\simeq5$ when
computing $p_\perp$-integrated yields (see below) in heavy-ion
collisions but not for the transverse energy $dE_t/dy$ or for
high-$p_t$ hadron production in p+p from fragmenting hard gluons.

\begin{figure}[htb]
\begin{center}
\includegraphics[width=10cm]{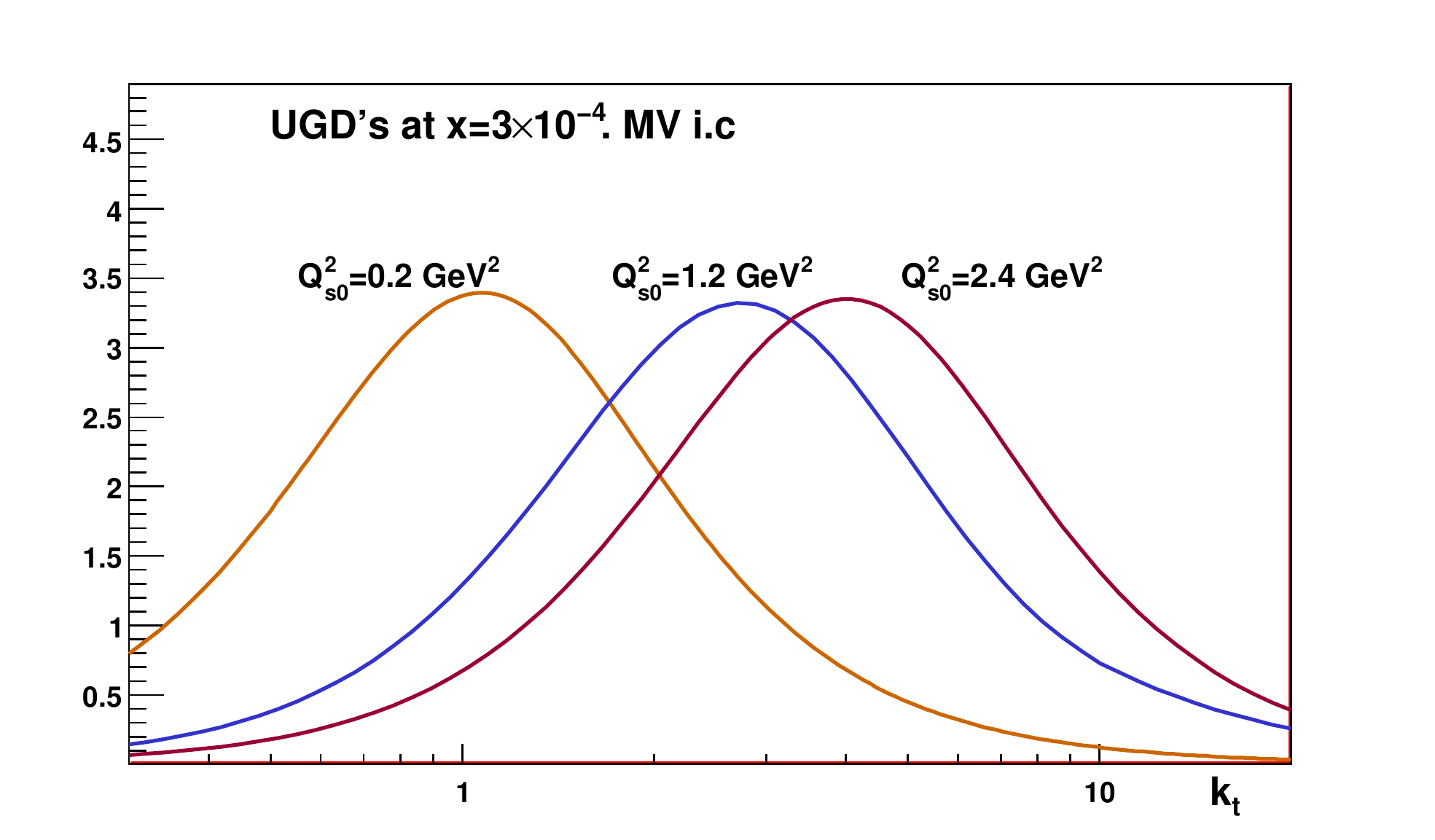}
\includegraphics[width=10cm]{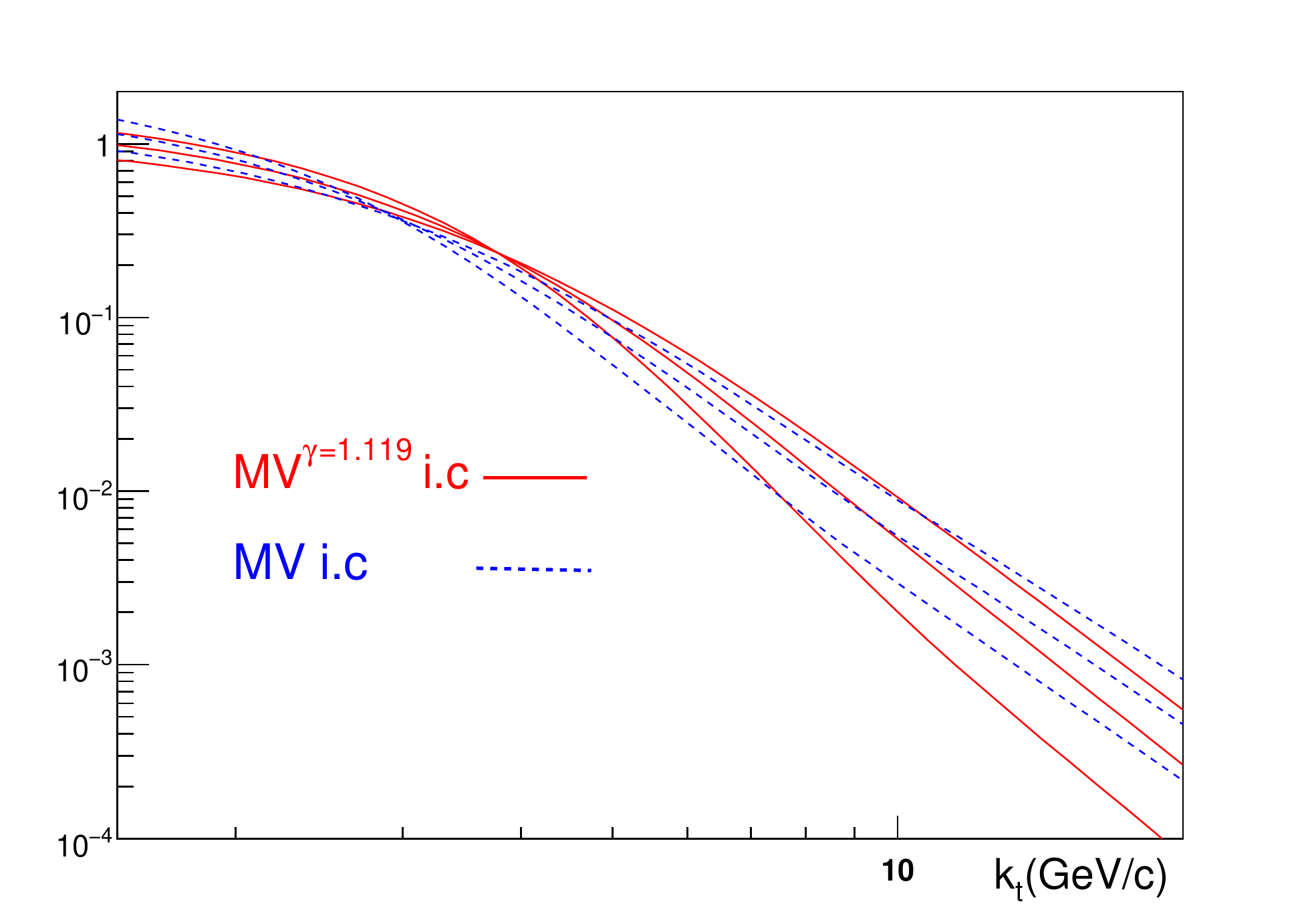}
\end{center}
\vspace*{-1cm}
\caption[a]{Unintegrated gluon distributions for different values of
  the initial saturation scale evolved to $x=3\cdot 10^{-4}$ (top) for MV i.c (top). UGD's for the two different initial conditions in the single nucleon case: MV$^{\gamma=1.119}$ (solid) and MV (dashed) at rapidities $Y=\ln(x_0/x)=1.5$, 3 and 6 (bottom). }
\label{ugds}
\end{figure}
In fig.~\ref{ugds} (top) we plot the ugd for three different initial
MV saturation scales at $x=3\cdot 10^{-4}$ versus transverse
momentum. The ugd corresponding to a single nucleon peaks at about
$k_t\simeq1$~GeV. The ugds for larger $Q_{s0}^2$ illustrate the shift
predicted for a 6-nucleon and 12-nucleon target, respectively. The bottom plot shows the different $k_t$ slopes for the two initial conditions (MV$^{\gamma=1.119}$ and MV) considered in this work in the single nucleon case.

\subsection{Observables}

Eq.~(\ref{kt2}) is the starting point for all observables shown below. In
particular, the charged particle multiplicity and the transverse
energy can be obtained by integrating over the transverse plane and
$p_t$,
\begin{eqnarray}
\frac{dN_{\rm ch}}{dy} &=& \frac{2}{3} \kappa_g \int d^2R \int d^2p_t \,
     \frac{dN^{A+B\to g}}{dy\, d^2p_t\, d^2R} \\
\frac{dN_{\rm ch}}{d^2p_t dy} &=&
\int d^2R \int \frac{dz}{z^2} \, D_h(z=\frac{p_t}{k_t}) \,
     \frac{dN^{A+B\to g}}{dy\, d^2k_t\, d^2R} \label{eq:FFconv} \\
\frac{dE_t}{dy} &=& \int d^2R \int d^2p_t \, p_t\,
     \frac{dN^{A+B\to g}}{dy\, d^2p_t\, d^2R} ~.
\end{eqnarray}
Note that a low-$p_t$ cutoff is not required since the integration
over $k_t$ in~(\ref{kt2}) extends only up to $p_t$. The saturation of
the gluon distribution functions guarantees that the dominant scale in
the transverse momentum integrations is the saturation momentum. For
the single-inclusive hadron $p_t$ distributions~(\ref{eq:FFconv}) we
use the KKP gluon $\to$ charged hadron LO fragmentation
function~\cite{KKP} with the scale $Q^2=k_t^2$; the integral over the
hadron momentum fraction is restricted to $z\ge0.05$ to avoid a
violation of the momentum sum rule. The ``gluon multiplication
factor'' is fixed to $\kappa_g=5$ (energy and centrality independent)
in order to reproduce the measured charged hadron multiplicity in
heavy-ion collisions at RHIC and LHC energies, see below. The upper
limit in the integrals over the gluon transverse momentum in $dN_{\rm
  ch}/dy$ and $dE_t/dy$ has been taken as $p_t^{\rm max}=12$~GeV; if
the integrals are extended further then a slight adjustment of the
normalization factors $\kappa_g$ and $K$ may be required.

\begin{figure}[htb]
\begin{center}
\includegraphics[width=10cm]{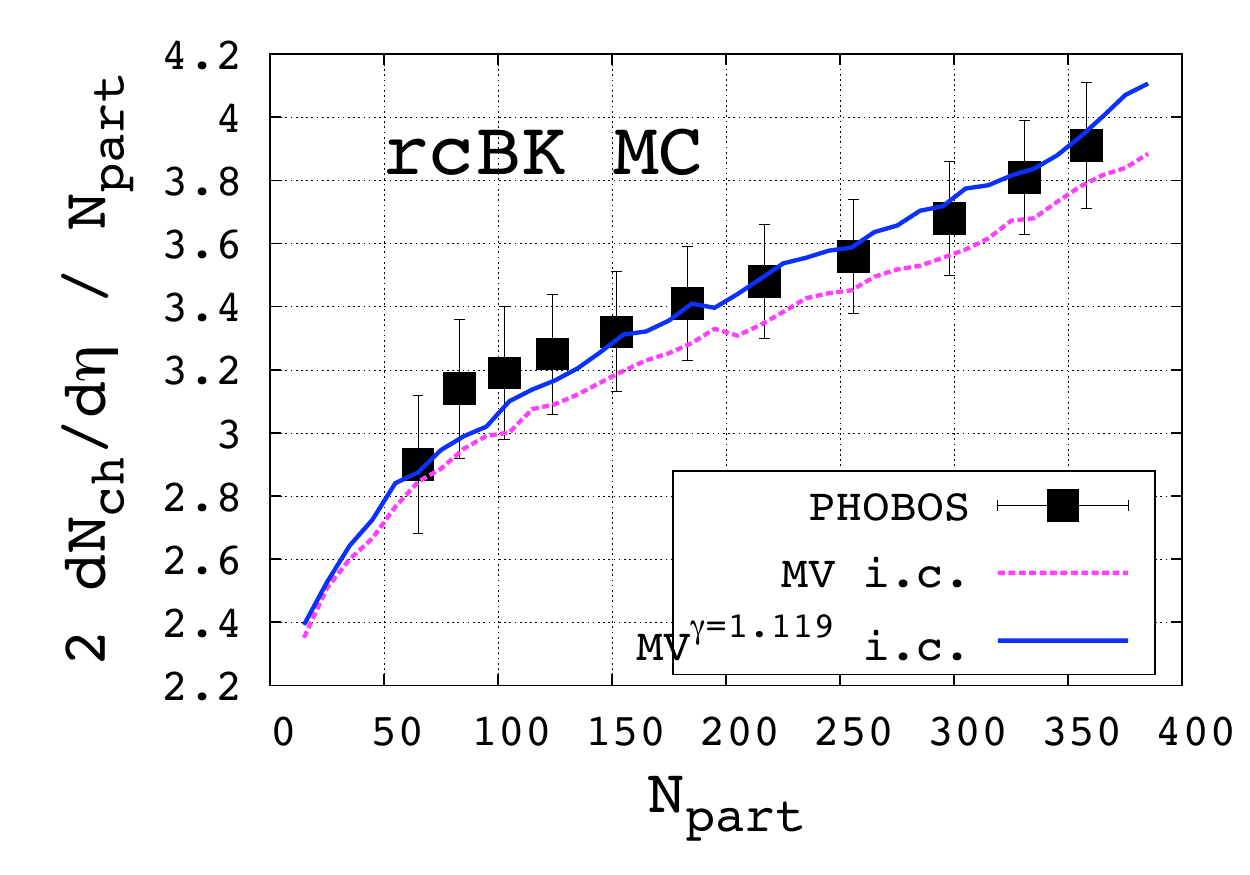}
\end{center}
\vspace*{-1cm}
\caption[a]{Centrality dependence of the charged particle multiplicity
  at midrapidity for Au+Au collisions at $\surd s = 200$~GeV. PHOBOS
  data: ref.~\protect\cite{Back:2002uc}.}
\label{fig:N_rhic}
\end{figure}

\begin{figure}[htb]
\begin{center}
\includegraphics[width=10cm]{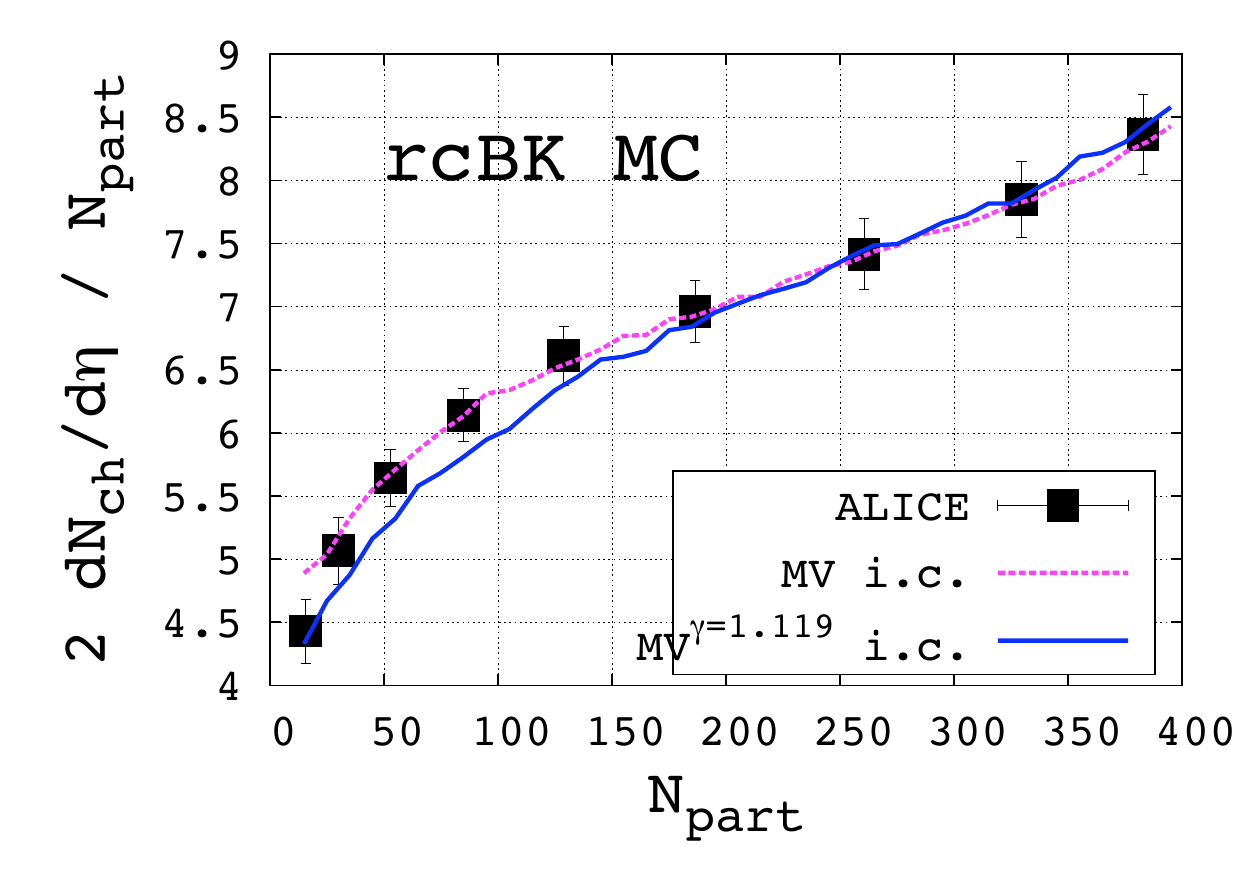}
\end{center}
\vspace*{-1cm}
\caption[a]{Centrality dependence of the charged particle multiplicity
  at midrapidity for Pb+Pb collisions at $\surd s = 2.76$~TeV. Alice data from
  ref.~\protect\cite{Aamodt:2010pb}.}
\label{fig:N_LHC}
\end{figure}

\begin{figure}[htb]
\begin{center}
\includegraphics[width=10cm]{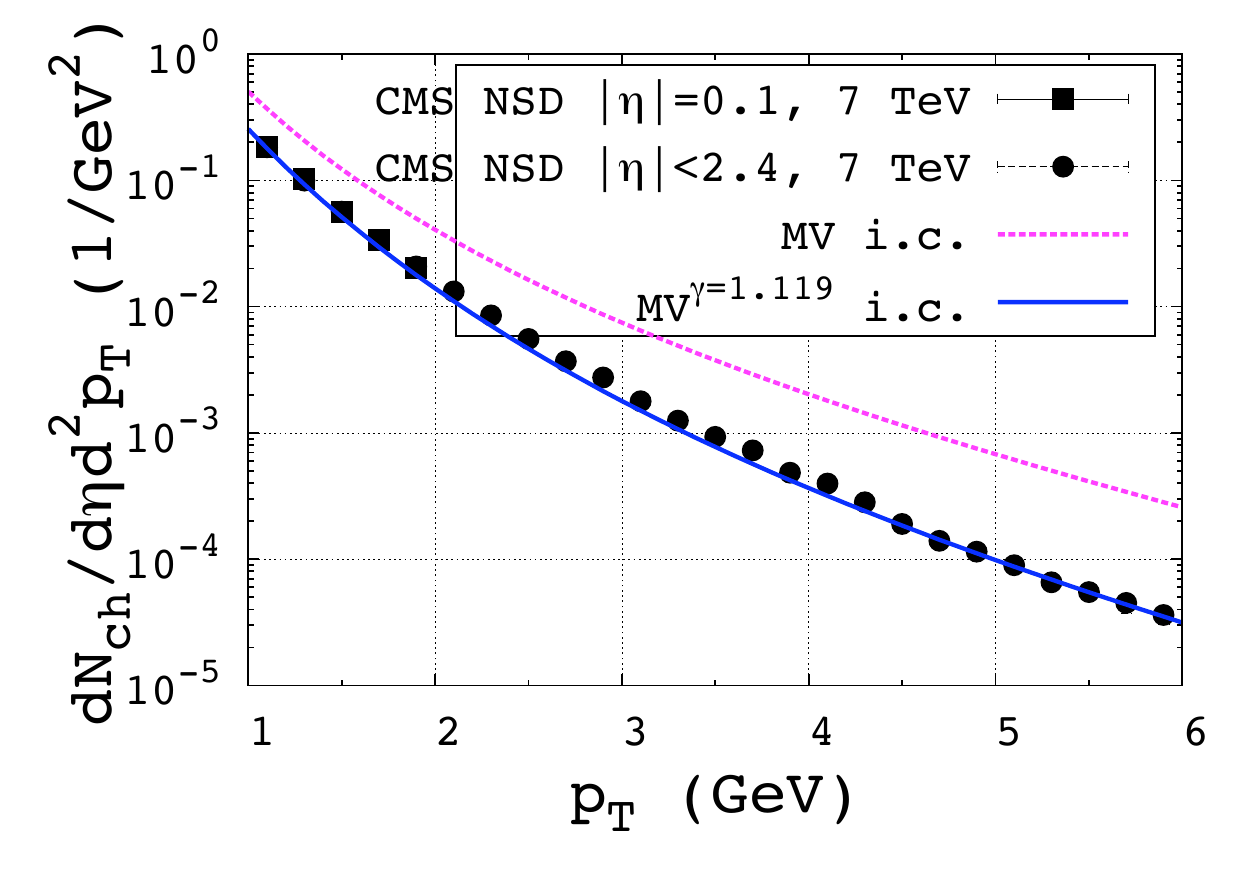}
\end{center}
\vspace*{-1cm}
\caption[a]{Transverse momentum distribution of charged particles
at $\eta=0$ for p+p collisions at $\surd s = 7$~TeV. CMS data from
ref.~\protect\cite{CMSpp7TeV}.}
\label{fig:pp7TeV}
\end{figure}

In order to compare our results for initial gluon production to the
final state distributions of detected particles one has to translate
the rapidity distributions into pseudo-rapidity distributions through
the $y\to\eta$ Jacobian,
\begin{eqnarray}
\frac{dN_{\rm ch}}{d\eta} &=& \frac{\cosh\eta}
{\sqrt{\cosh^2\eta + m^2/P^2}}\, \frac{dN_{\rm ch}}{dy} \\
\frac{dE_t}{d\eta} &=& \frac{\cosh\eta}
{\sqrt{\cosh^2\eta + m^2/P^2}}\, \frac{dE_t}{dy}~,
\end{eqnarray}
with $y=\frac{1}{2}\ln\, (\sqrt{\cosh^2\eta + m^2/P^2}+\sinh\eta)/
(\sqrt{\cosh^2\eta + m^2/P^2}-\sinh\eta)$. We
assume that in this Jacobian $m=350$~MeV and $P=0.13~\mathrm{GeV} + 
0.32~\mathrm{GeV} \sqrt{s/1~\mathrm{TeV}}^{\, 0.115}$.
Note that such transformation is not needed (it is trivial) if
one is interested in initial (massless) gluon production to initialize
a hydrodynamic simulation.

\section{Results}
In this section we present some results obtained with the new
MV$^{\gamma=1.119}$ uGD and compare them to the original uGD obtained by rcBK
evolution with MV model initial conditions.

In Fig.~\ref{fig:N_rhic} we show the centrality dependence of $dN/d\eta$
at full RHIC energy. This energy probes mainly the initial condition
for small-$x$ evolution. Both sets perform reasonably well and we do
not find it possible to distinguish.

Fig.~\ref{fig:N_LHC} shows the centrality dependence of $dN/d\eta$ at
$\eta=0$ for Pb+Pb collisions at $\surd s = 2.76$~TeV \footnote{The
  Alice papers summarize other recent predictions for $dN_{ch}/d\eta$
  for Pb+Pb collisions.}. Once again both uGD sets agree with the data
to within a few percent. It is worth noting that our present results
for the multiplicity in central collisions are higher than the one
predicted in ref.~\cite{Albacete:2007sm}, which also relied on the
rcBK equation plus $k_t$-factorization. The main difference between
these two works relates to the treatment of the geometry of the
collision: while in \cite{Albacete:2007sm} the full nucleus was
characterized by a single average saturation scale and then evolved to
higher energies, the Monte Carlo approach presented here allows for
different local densities at every point in the transverse plane, each
of them evolved locally to higher energies. The average over different
configurations is performed after the evolution, and not before, as
implicitly done in \cite{Albacete:2007sm}. Thus we interpret these
two different results as an indication that the average over nuclear
geometry does not commute with the evolution.

In Fig.~\ref{fig:pp7TeV} we show the transverse momentum distribution
of charged particles for p+p collisions at $\surd s = 7$~TeV. We
recall that presently we assume uniform and homogeneous,
``disc-like'' nucleons. However, we do not attempt to compute the
inelastic proton-proton cross section and high-$p_t$ hadron spectra
may not be very sensitive to the proton impact parameter profile.

For the range of $p_\perp$ shown in the figure, particle production
probes LC momentum fractions well below our assumed starting
point of $x_0=0.01$. The uGD derived from MV model initial conditions
is clearly too ``hard'' and predicts an incorrect slope. The new uGD
obtained from the MV$^{\gamma=1.119}$ initial condition corrects this
deficiency and provides a good description of the CMS data in the
small-$x$, semi-hard regime (see, also,
ref.~\cite{Tribedy:2010ab}). This illustrates the power of LHC to
constrain small-$x$ physics. Also, we have used this observable to fix
the genuine ``K-factor'' to $K=2$ (MV$^{\gamma=1.119}$ i.c.) or $K=1.5$ (MV
i.c.), respectively.

\begin{figure}[htb]
\begin{center}
\includegraphics[width=8.5cm]{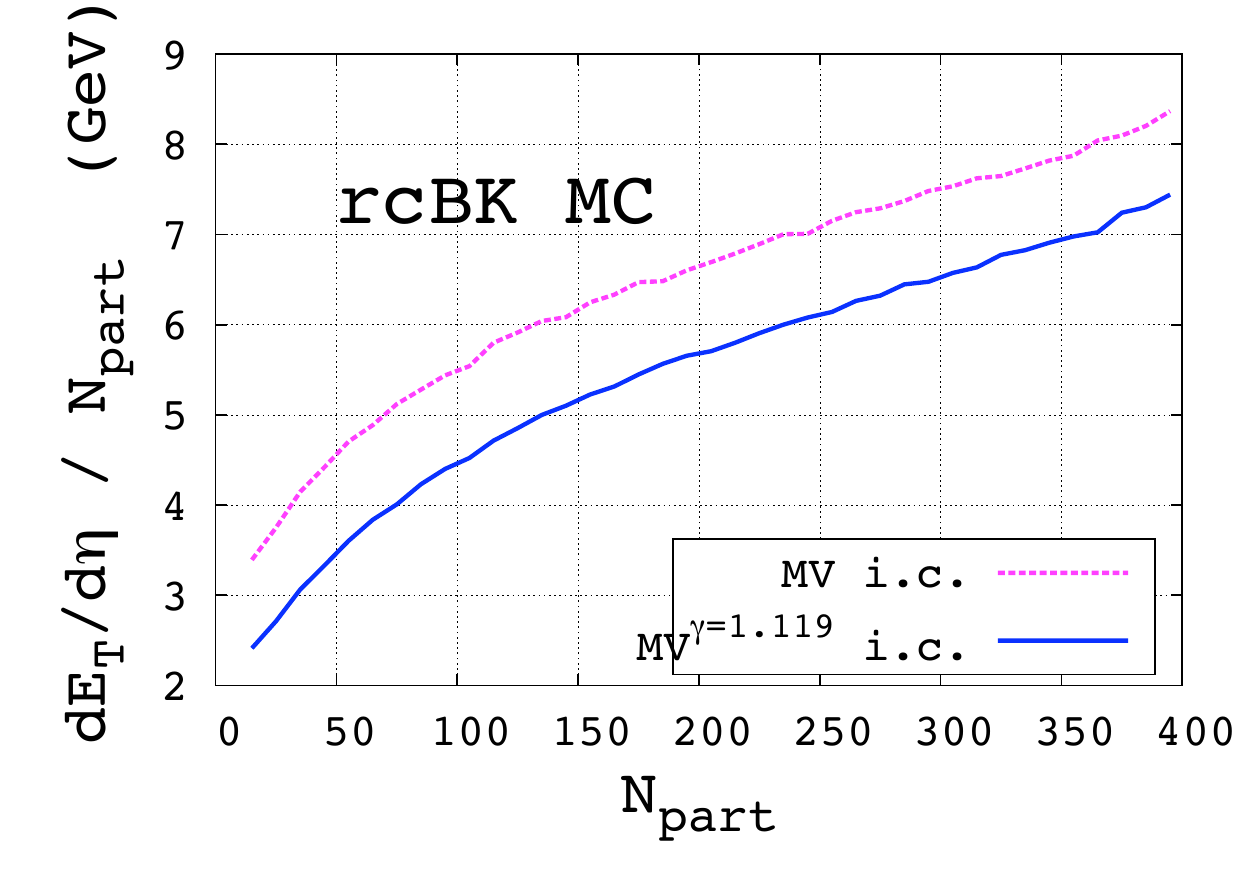}
\includegraphics[width=8.5cm]{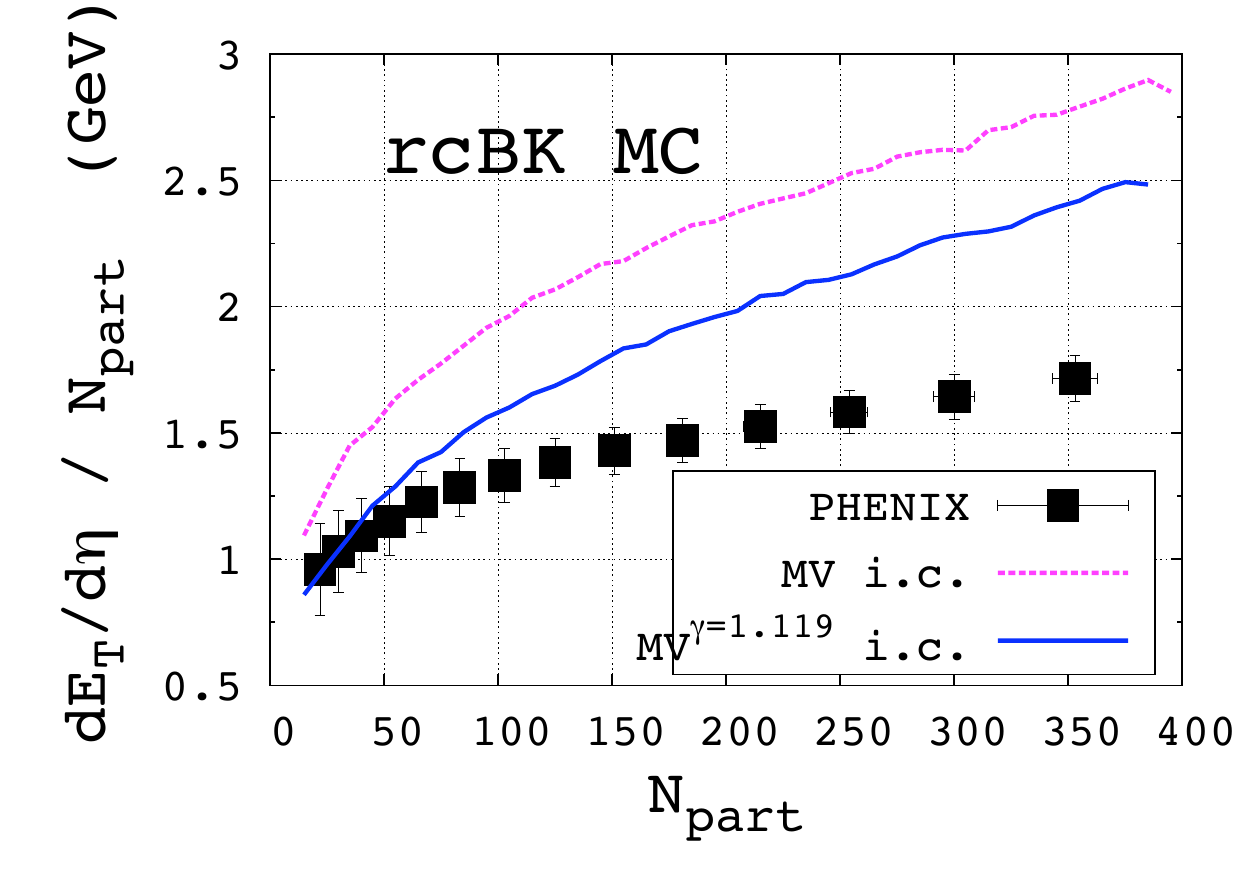}
\end{center}
\vspace*{-1cm}
\caption[a]{Transverse energy at $\eta=0$ versus centrality
  for Pb+Pb collisions at $\surd s = 2.76$~TeV (left) and for Au+Au
  collisions at $\surd s = 200$~GeV (right).}
\label{fig:Et}
\end{figure}
The transverse energy in heavy-ion collisions versus centrality is
shown in Fig.~\ref{fig:Et}. The new MV$^{\gamma=1.119}$ uGD predicts a lower
transverse energy than the one derived from MV model initial
conditions, by about 1~GeV per participant at LHC
energies\footnote{Note that we deviate from v1 of this manuscript in
  that the ``gluon multiplication factor'' $\kappa_g$ has been dropped
  from the normalization for $E_\perp$.}. It should be kept in mind
that the measured transverse energy in the final state is only a lower
bound on the initial $dE_\perp/d\eta$ since soft final-state
interactions such as collective longitudinal flow do not conserve the
transverse energy per unit of rapidity~\cite{dEt_red}.

\begin{figure}[htb]
\begin{center}
\includegraphics[width=8.5cm]{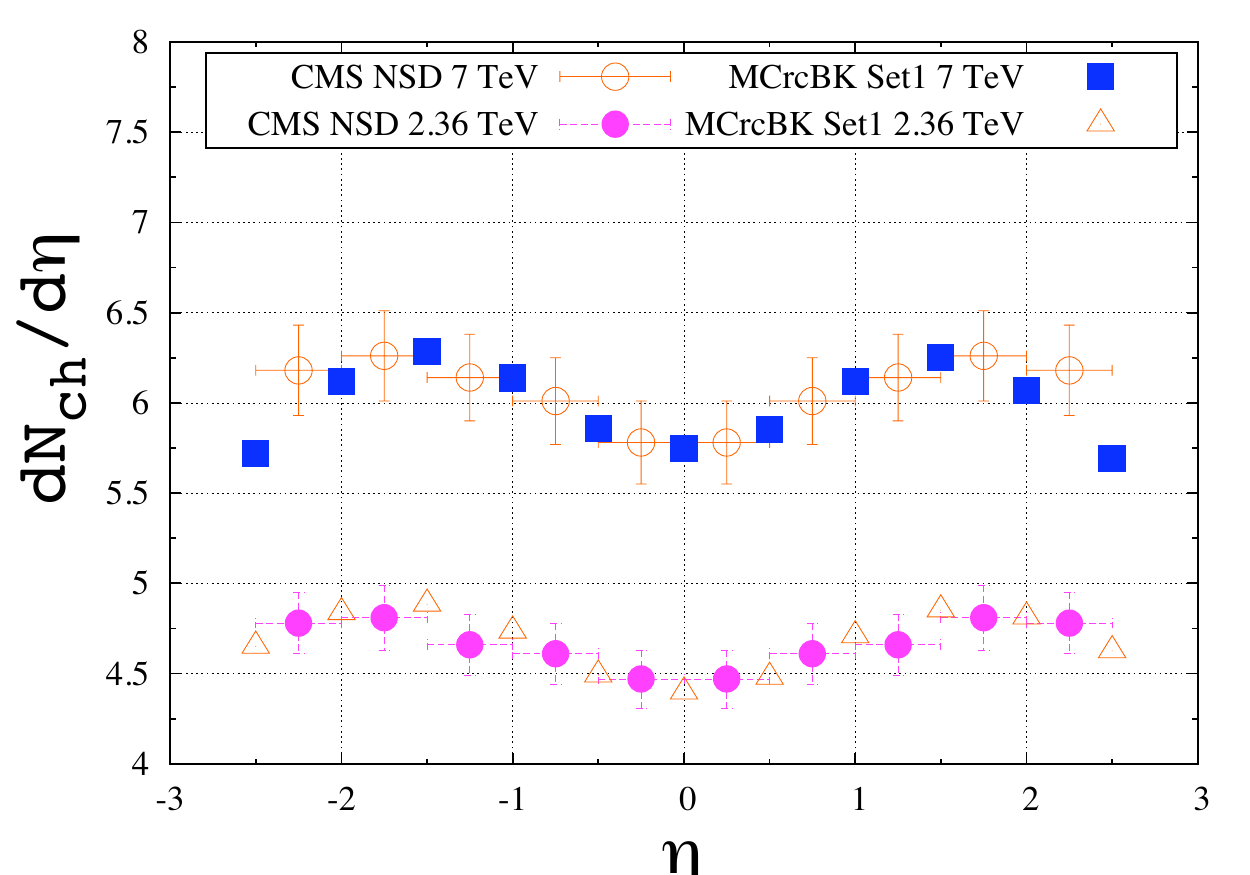}
\includegraphics[width=8.5cm]{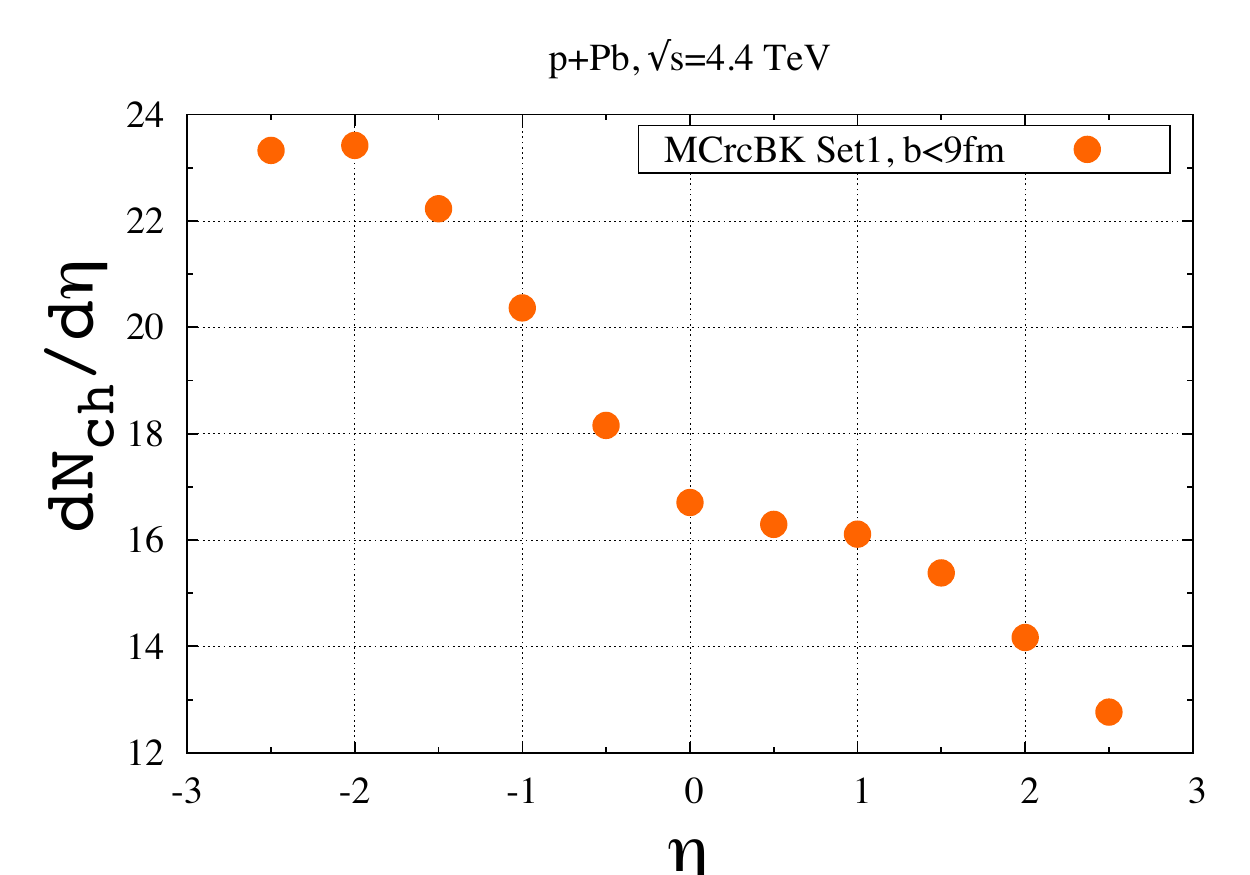}
\end{center}
\caption[a]{Charged particle pseudo-rapidity distributions for UGD Set
  1 (MV model initial conditions) in pp collisions at 7~TeV
  (normalization adjustement -10\%), 2.36~TeV (normalization
  adjustement -4\%), and p+Pb collisions at 4.4~TeV (standard
  normalization as for AA collisions from Figs.~\ref{fig:N_rhic},
  \ref{fig:N_LHC}).
}
\label{fig:dNdeta}
\end{figure}
In Fig.\ref{fig:dNdeta} we show the rapidity distributions in pp
collisions at two LHC energies and a prediction for minimum bias p+Pb
collisions at 4.4~TeV. We restrict these distributions to
$|\eta|\le2.5$ since we have not accounted for the contribution of
scattered valence partons. While there is perhaps an overall
normalization uncertainty of up to $\sim 10\%$, we note the large
forward-backward asymmetry for p+Pb collisions\footnote{See also discussion
  of asymmetric collisions in ref.~\cite{Kharzeev:2002ei}.}at
rapidities far from the valence partons predicted by the rcBK-UGD.

We leave a more detailed study of the sensitivity of our results to
the parameters of the model and a more detailed comparison with RHIC
and LHC data for a future publication.

\end{document}